\documentclass[aps,twocolumn,prc,groupedaddress,longbibliography]{revtex4-2}
\usepackage{amsmath}
\usepackage{graphicx}
\usepackage{color}
\usepackage{here}
\usepackage{bm}
\usepackage{subfigure}

\begin{document}

\title{Screening of nucleon electric dipole moments in atomic systems}

\author{Kota~Yanase}
\email{yanase@cns.s.u-tokyo.ac.jp}
\affiliation{Center for Nuclear Study, The University of Tokyo, Hongo Tokyo 113-0033, Japan}

\date{\today}

\begin{abstract}
The electric dipole moments (EDMs) of diamagnetic atoms are expected to be sensitive to charge-parity violation particularly in nuclei through the nuclear Schiff moment.
I explicitly demonstrate that the well-known form of the Schiff moment operator originating from the nucleon EDM is obtained by considering the screening of the nucleon EDMs in a neutral atom.
Consequently, an additional contribution to the Schiff moment arises from the screening of the nuclear EDM induced by the interaction of the nucleon EDMs with the protons.
This correction to the Schiff moment of $^{199}$Hg is evaluated in the independent particle model.
\end{abstract}

\maketitle

\section{Schiff moment \label{sec: S1}}

The observation of a permanent electric dipole moment (EDM) of an atom implies the existence of parity ($ P $) and time-reversal ($ T $) violating interactions between constituent particles.
The atomic EDM is defined by
\begin{align}
 \bm{ d }_{ \text{atom} }
 =
 -
 \sum_{ i = 1 }^{ Z }
 e \bm{ r }_{ i }'
 ,
\end{align}
where $ e $ is the elementary charge and $ \bm{ r }_{ i }' $ indicates the coordinates of the atomic electrons.
The interaction of the atomic EDM with an external electric field causes an energy shift to be measured~\cite{Khriplovich-Lamoreaux-CP,Chupp2019-review}.

Although the nuclear EDM and the intrinsic EDMs of electrons and nucleons are independently coupled to the external electric field $ \bm{ E }_{ \text{ext} } $ as
\begin{align}
 V_{ \text{ext} }
 =
 -
 \left[
  \bm{ d }_{ \text{atom} }
  +
  \bm{ d }_{ \text{nucl} }
  +
  \sum_{ i = 1 }^{ Z }
  \bm{ d }^{( e )}_{ i }
  +
  \sum_{ a = 1 }^{ A }
  \bm{ d }_{ a }
 \right]
 \! \cdot \!
 \bm{ E }_{ \text{ext} }
 \label{eq: V_ext}
 ,
\end{align}
they are obscured by the internal interactions with the electrons.
Here $ \bm{ d }^{( e )}_{ i } $ and $ \bm{ d }_{ a } $ are the intrinsic EDMs of electrons and nucleons, respectively, and the nuclear EDM is defined by
\begin{align}
 \bm{ d }_{ \text{nucl} }
 =
 \sum_{ a = 1 }^{ Z }
 e \bm{ r }_{ a }
 \label{eq: nuclear EDM}
 ,
\end{align}
where $ \bm{ r }_{ a } $ indicates the proton coordinates.
In particular, the energy shift due to the EDM of a point-like nucleus is canceled by the contribution from the atomic EDM induced by the internal interaction of the nuclear EDM with the electrons.
However, $ P $, $ T $-odd nucleon-nucleon ($ NN $) interactions allow a finite-size nucleus to have the nuclear Schiff moment as well as the nuclear EDM.
Since the atomic EDM induced by the interaction of the Schiff moment with the electrons survives the screening, the atomic EDMs particularly of diamagnetic atoms are sensitive to $ P $, $ T $-odd $ NN $ interactions~\cite{Schiff1963,Sushkov1984,Spevak1997,Liu2007-NSM}.
The screening mechanism of the nuclear EDM induced by the $ P $, $ T $-odd meson-exchange $ NN $ ($ \pi NN $) interaction is reviewed in this section.

The Hamiltonian of an atomic system that conserves $ P $ and $ T $ symmetries is written as
\begin{align}
 &
 H_{ \text{atom} }
 =
 H_{ \text{nucl} }
 +
 H_{ e }
 \label{eq: H_atom}
 ,
 \\
 &
 H_{ e }
 =
 T_{ e }
 + V^{( ee )}
 + V_{ \text{even}-l }^{( eN )}
 \label{eq: H_e}
 ,
\end{align}
where $ H_{ \text{nucl} } $ denotes $ P $, $ T $-even $ NN $ interactions.
The electron kinetic term $ T_{ e } $ and the electron-electron interactions $ V^{( ee )} $ are not relevant to the nuclear $ P $, $ T $ violation of interest.
The electrostatic interaction between the electrons and the protons is
\begin{align}
 &
 V^{( eN )}
 =
 - e^{ 2 }
 \sum_{ i = 1 }^{ Z }
 \sum_{ a = 1 }^{ Z }
 \frac{ 1 }{ \left| \bm{ r }_{ i }' - \bm{ r }_{ a } \right| }
 \label{eq: charge int}
 .
\end{align}
If $ r_{ i }' > r_{ a } $, then each term can be expanded as
\begin{align}
 \frac{ 1 }{ \left| \bm{ r }_{ i }' - \bm{ r }_{ a } \right| }
 & =
 \sum_{ l = 0 }^{ \infty }
 \frac{ ( -1 )^{ l } }{ l ! }
 \big( \bm{ r }_{ a } \! \cdot \! \bm{ \nabla }_{ i }' \big)^{ l }
 \frac{ 1 }{ r_{ i }' }
 \label{eq: multipole exp of charge int}
 .
\end{align}
The atomic Hamiltonian $ H_{ \text{atom} } $ does not contain the odd-$ l $ electron-nucleon ($ eN $) interactions denoted by $ V^{( eN )}_{ \text{odd}-l } $, which vanish unless $ P $ and $ T $ symmetries are both violated in the nucleus.

\begin{figure*}[htb]
\begin{center}
\subfigure[]{
 \includegraphics[height=4.0cm]{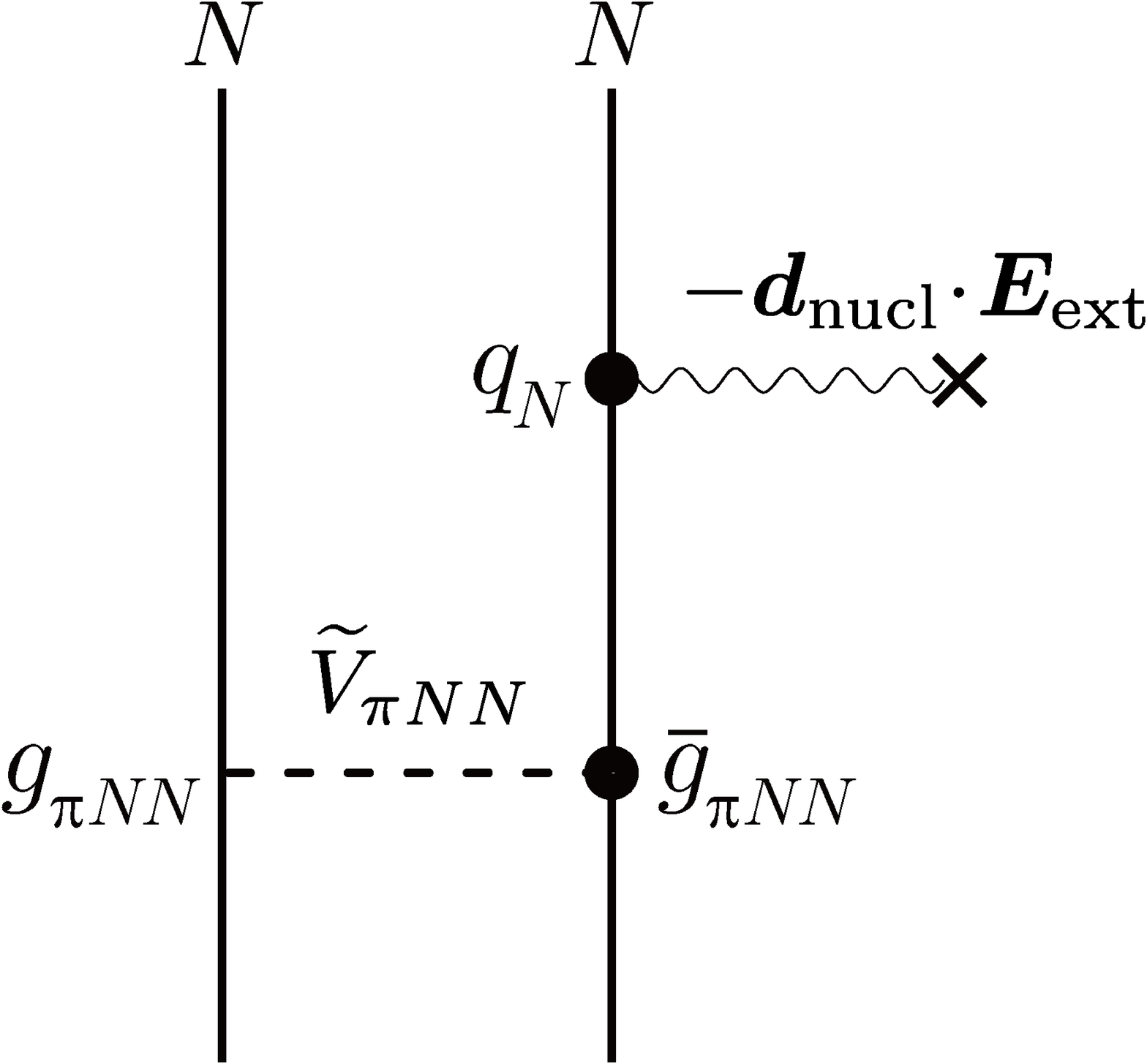}
\label{diagram: Schiff moment, second order}
}
\hspace{10pt}
\subfigure[]{
 \includegraphics[height=4.0cm]{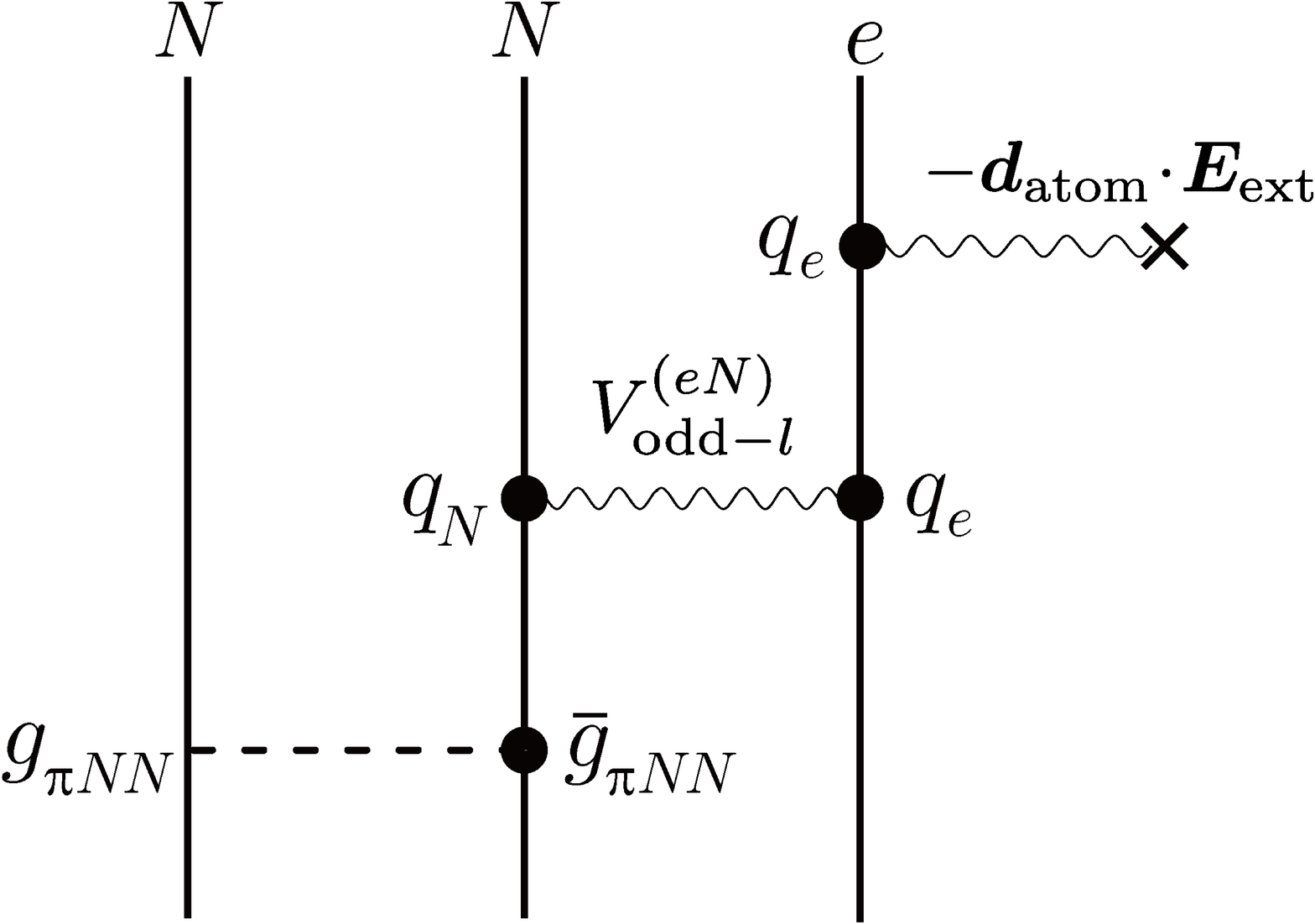}
\label{diagram: Schiff moment, third order}
}
\caption{
\label{diagram: Schiff moment}
(a) The second-order and (b) the third-order contributions of the $ P $, $ T $-odd $ \pi NN $ interaction $ \widetilde{ V }_{ \pi NN } $ to the energy shift of an atom immersed in an external electric field.
The black circles represent the $ P $, $ T $-odd vertices.
The electric charges of proton and electron are denoted by $ q_{ N } $ and $ q_{ e } $, respectively.
The $ P $, $ T $-odd $ \pi NN $ coupling constant is denoted by $ \overline{ g }_{ \pi NN } $.
}
\end{center}
\end{figure*}

The nuclear ground state in the existence of the $ P $, $ T $-odd $ \pi NN $ interaction $ \widetilde{ V }_{ \pi NN } $ is given by
\begin{align}
 &
 \big| \widetilde{ \psi }^{( N )}_{ \text{g.s.} } \big\rangle
 =
 \big| \psi^{( N )}_{ \text{g.s.} } \big\rangle
 \notag\\
 & \quad
 +
 \sum_{ n }
 \frac{ 1 }{ E^{( N )}_{ \text{g.s.} } - E^{( N )}_{ n } }
 \big| \psi^{( N )}_{ n } \big\rangle
 \big\langle \psi^{( N )}_{ n } \big|
  \widetilde{ V }_{ \pi NN }
 \big| \psi^{( N )}_{ \text{g.s.} } \big\rangle
 \label{PTV nuclear wave function}
 ,
\end{align}
where $ E^{( N )}_{ \text{g.s.} } $ and $ E^{( N )}_{ n } $ denote the energies of the ground state $ \big| \psi^{( N )}_{ \text{g.s.} } \big\rangle $ and excited states $ \big| \psi^{( N )}_{ n } \big\rangle $ of the nuclear Hamiltonian $ H_{ \text{nucl} } $, respectively.
As well as the atomic EDM is generated by $ P $, $ T $-violations in the electron system, the $ P $, $ T $-odd $ \pi NN $ interaction can induce the nuclear EDM.
The external interaction of the nuclear EDM represented in Fig.~\ref{diagram: Schiff moment, second order} causes the energy shift
\begin{align}
 \Delta E_{ 2 }
 ( \overline{ g }_{ \pi NN }, q_{ N } )
 =
 \big\langle \widetilde{ \psi }^{( N )}_{ \text{g.s.} } \big|
  -
  \bm{ d }_{ \text{nucl} } \! \cdot \! \bm{ E }_{ \text{ext} }
 \big| \widetilde{ \psi }^{( N )}_{ \text{g.s.} } \big\rangle
 \label{eq: Delta E2 ( gn, qn )}
 ,
\end{align}
where the coupling constants $ \overline{ g }_{ \pi NN } $ and $ q_{ N } $ specify the perturbative interactions.


The $ P $, $ T $-odd $ \pi NN $ interaction also induces the odd-$ l $ $ eN $ interactions $ V^{( eN )}_{ \text{odd}-l } $, which violate $ P $ and $ T $ symmetries in the electron system.
Consequently, the atomic EDM contributes to the energy shift in third order perturbation as
\begin{align}
 &
 \Delta E_{ 3 }
 ( \overline{ g }_{ \pi NN }, q_{ N }, q_{ e } )
 =
 \sum_{ m }
 \frac{ 1 }{ E_{ \text{g.s.} }^{( e )} - E_{ m }^{( e )} }
 \notag\\
 & \quad \times
 \big\langle \widetilde{ \psi }^{( A )}_{ \text{g.s.} } \big|
  - \bm{ d }_{ \text{atom} } \! \cdot \! \bm{ E }_{ \text{ext} }
 \big| \widetilde{ \psi }^{( A )}_{ m } \big\rangle
 \big\langle \widetilde{ \psi }^{( A )}_{ m } \big|
  V^{( eN )}_{ \text{odd} - l }
 \big| \widetilde{ \psi }^{( A )}_{ \text{g.s.} } \big\rangle
 \notag\\
 & \quad
 + c.c.
 \label{eq: Delta E3 ( gn, qn, qe )}
 \end{align}
This process is represented in Fig.~\ref{diagram: Schiff moment, third order}.
The eigenstates of the atomic system in the existence of the $ P $, $ T $-odd $ \pi NN $ interaction are expressed except for the Clebsch-Gordan coefficients as
\begin{align}
 &
 \big| \widetilde{ \psi }^{( A )}_{ \text{g.s.} } \big\rangle
 =
 \big| \widetilde{ \psi }^{( N )}_{ \text{g.s.} } \big\rangle
 \otimes
 \big| \psi^{( e )}_{ \text{g.s.} } \big\rangle
 ,
 \\
 &
 \big| \widetilde{ \psi }^{( A )}_{ m } \big\rangle
 =
 \big| \widetilde{ \psi }^{( N )}_{ \text{g.s.} } \big\rangle
 \otimes
 \big| \psi^{( e )}_{ m } \big\rangle
 ,
\end{align}
where $ \big| \psi^{( e )}_{ \text{g.s.} } \big\rangle $ and $ \big| \psi^{( e )}_{ m } \big\rangle $ denote the ground state and excited states of the electron system described by $ H_{ e } $ with the energies $ E^{( e )}_{ \text{g.s.} } $ and $ E^{( e )}_{ m } $, respectively.

Here I summarize the notations used in this paper.
The superscripts $ ( A ) $, $ ( N ) $, and $ ( e ) $ represent the atomic, nuclear, and electron systems, respectively.
$ P $, $ T $-odd interactions and $ P $, $ T $-violated wave functions are denoted by $ \widetilde{ V } $ and $ \big| \widetilde{ \psi } \big\rangle $, respectively.

The screening of the nuclear EDM is demonstrated by using a Hermitian operator~\cite{Senkov2008}
\begin{align}
 U_{ \text{nucl} }
 =
 i \frac{ 1 }{ Ze }
 \left\langle \bm{ d }_{ \text{nucl} } \right\rangle
 \! \cdot \!
 \sum_{ i = 1 }^{ Z } \bm{ \nabla }_{ i }'
 \label{eq: Hermitian operator for nuclear EDM}
 ,
\end{align}
where
\begin{align}
 \left\langle \bm{ d }_{ \text{nucl} } \right\rangle
 =
 \big\langle \widetilde{ \psi }^{( N )}_{ \text{g.s.} } \big|
  \bm{ d }_{ \text{nucl} }
 \big| \widetilde{ \psi }^{( N )}_{ \text{g.s.} } \big\rangle
 .
\end{align}
The nuclear EDM interactions in Eqs.~(\ref{eq: Delta E2 ( gn, qn )}) and (\ref{eq: Delta E3 ( gn, qn, qe )}) are transformed as
\begin{align}
 &
 \big\langle \widetilde{ \psi }^{( A )}_{ \text{g.s.} } \big|
  -
  \bm{ d }_{ \text{nucl} } \! \cdot \! \bm{ E }_{ \text{ext} }
 \big| \widetilde{ \psi }^{( A )}_{ \text{g.s.} } \big\rangle
 \notag\\
 & \quad
 =
 i
 \big\langle \widetilde{ \psi }^{( A )}_{ \text{g.s.} } \big|
  \big[
   U_{ \text{nucl} },
   - \bm{ d }_{ \text{atom} } \! \cdot \! \bm{ E }_{ \text{ext} }
  \big]
 \big| \widetilde{ \psi }^{( A )}_{ \text{g.s.} } \big\rangle
 \label{eq: U_N trans of external int of nucleus}
 ,
\end{align}
and
\begin{align}
 &
 \big\langle \widetilde{ \psi }^{( A )}_{ m } \big|
  V^{( eN )}_{ l = 1 }
 \big| \widetilde{ \psi }^{( A )}_{ \text{g.s.} } \big\rangle
 \notag\\
 & \quad
 =
 i
 \big\langle \widetilde{ \psi }^{( A )}_{ m } \big|
  \big[
   U_{ \text{nucl} } , V^{( eN )}_{ l = 0 }
  \big]
 \big| \widetilde{ \psi }^{( A )}_{ \text{g.s.} } \big\rangle
 \notag\\
 & \quad
 =
 i
 \big\langle \widetilde{ \psi }^{( A )}_{ m } \big|
  \big[
   U_{ \text{nucl} } , H_{ e }
  \big]
 \big| \widetilde{ \psi }^{( A )}_{ \text{g.s.} } \big\rangle
 \notag\\
 & \qquad
 - i
 \big\langle \widetilde{ \psi }^{( A )}_{ m } \big|
  \big[
   U_{ \text{nucl} },
   V^{( eN )}_{ l = 2 }
   +
   V^{( eN )}_{ l = 4 }
   + \cdots
  \big]
 \big| \widetilde{ \psi }^{( A )}_{ \text{g.s.} } \big\rangle
 \label{eq: U_N trans of charge int of nucleus, l = 1}
 .
\end{align}
The $ eN $ interactions of a point-like nucleus consist of the $ l = 0, 1 $ components, which are explicitly given by
\begin{align}
 V^{( eN )}_{ l = 0 }
 & =
 - Z e^{ 2 }
 \sum_{ i = 1 }^{ Z }
 \frac{ 1 }{ r_{ i }' }
 ,
 \\
 V^{( eN )}_{ l = 1 }
 & =
 e \bm{ d }_{ \text{nucl} }
 \! \cdot \!
 \sum_{ i = 1 }^{ Z }
 \bm{ \nabla }_{ i }'
 \frac{ 1 }{ r_{ i }' }
 .
\end{align}
The last equality in Eq.~(\ref{eq: U_N trans of charge int of nucleus, l = 1}) follows from the fact that the operator $ U_{ \text{nucl} } $ commutes with the electron kinetic term $ T_{ e } $ and the interactions between electrons $ V^{( ee )} $.

Although the same transformations are realized even if one adopts
\begin{align}
 U_{ \text{nucl} }'
 =
 i \frac{ 1 }{ Ze }
 \bm{ d }_{ \text{nucl} }
 \! \cdot \!
 \sum_{ i = 1 }^{ Z } \bm{ \nabla }_{ i }'
\end{align}
instead of $ U_{ \text{nucl} } $, the resulting nuclear moment is a more complicated two-body operator than the Schiff moment~(\ref{eq: S1 operator}).

Using the transformations (\ref{eq: U_N trans of external int of nucleus}) and (\ref{eq: U_N trans of charge int of nucleus, l = 1}), the third-order effect~(\ref{eq: Delta E3 ( gn, qn, qe )}) is transformed as
\begin{align}
 &
 \Delta E_{ 3 }
 ( \overline{ g }_{ \pi NN }, q_{ N }, q_{ e } )
 =
 -
 \Delta E_{ 2 }
 ( \overline{ g }_{ \pi NN }, q_{ N } )
 \notag\\
 & \quad
 +
 \sum_{ m }
 \frac{ 1 }{ E^{( e )}_{ \text{g.s.} } - E^{( e )}_{ m } }
 \notag\\
 & \qquad \times
 \Big[
 \big\langle \psi^{( e )}_{ \text{g.s.} } \big|
  - \bm{ d }_{ \text{atom} } \! \cdot \! \bm{ E }_{ \text{ext} }
 \big| \psi^{( e )}_{ m } \big\rangle
 \big\langle \psi^{( e )}_{ m } \big|
  V_{ \text{NSM-1} }
 \big| \psi^{( e )}_{ \text{g.s.} } \big\rangle
 \notag\\
 & \qquad \qquad
 + c.c.
 \Big]
 \label{3rd-order effect of piNN, Schiff moment}
 ,
\end{align}
where the first term implies the screening of the nuclear EDM.
The remaining terms caused by the finite-size effect can be nonzero in the ``point-like nucleus limit'', where
\begin{align}
 \bm{ \nabla }_{ i }'^{ 2 }
 \frac{ 1 }{ r_{ i }' }
 \Big|_{ R \rightarrow 0 }
 =
 - 4 \pi \delta ( \bm{ r }_{ i }' )
 .
\end{align}
Here $ R $ is the nuclear radius.

Considering $ l \leq 3 $, one obtains
\begin{align}
 &
 \big\langle \psi^{( e )}_{ m } \big|
  V_{ \text{NSM-1} }
 \big| \psi^{( e )}_{ \text{g.s.} } \big\rangle
 \notag\\
 & \quad
 =
 \big\langle \widetilde{ \psi }^{( A )}_{ m } \big|
  V^{( eN )}_{ l = 3 }
 \big| \widetilde{ \psi }^{( A )}_{ \text{g.s.} } \big\rangle
 \notag\\
 & \qquad
 - i
 \big\langle \widetilde{ \psi }^{( A )}_{ m } \big|
  \big[
   U_{ \text{nucl} } , V^{( eN )}_{ l = 2 }
  \big]
 \big| \widetilde{ \psi }^{( A )}_{ \text{g.s.} } \big\rangle
 \notag\\
 & \quad
 =
 \big\langle \psi^{( e )}_{ m } \big|
  - 4 \pi e
  \sum_{ i = 1 }^{ Z }
  \big\langle \bm{ S }_{ 1 } \big\rangle
  \! \cdot \!
  \bm{ \nabla }_{ i }' \delta ( \bm{ r }_{ i }' )
 \big| \psi^{( e )}_{ \text{g.s.} } \big\rangle
 \label{eq: NSM-1 int}
 ,
\end{align}
where the nuclear part is separated from the electron part as explained in Appendix~\ref{app: Schiff from pi-NN}.
The expectation value of the nuclear Schiff moment is given by
\begin{align}
 \big\langle \bm{ S }_{ 1 } \big\rangle
 =
 &
 \sum_{ n }
 \frac{ 1 }{ E^{( N )}_{ \text{g.s.} } - E^{( N )}_{ n } }
 \notag\\
 & \qquad \times
 \big\langle \psi^{( N )}_{ \text{g.s.} } \big|
  \bm{ S }_{ 1 }
 \big| \psi^{( N )}_{ n } \big\rangle
 \big\langle \psi^{( N )}_{ n } \big|
  \widetilde{ V }_{ \pi NN }
 \big| \psi^{( N )}_{ \text{g.s.} } \big\rangle
 \notag\\
 &
 + c.c.
\end{align}
The explicit form of the Schiff moment operator is
\begin{align}
 S_{ 1, k }
 =
 \frac{ e }{ 10 }
 \sum_{ a = 1 }^{ Z }
 \bigg[
  r_{ a }^{ 2 }
  r_{ a, k }
  -
  \frac{ 5 }{ 3 }
  r_{ a, k }
  \left\langle r^{ 2 } \right\rangle_{ \text{ch} }
  -
  \frac{ 4 }{ 3 }
  r_{ a, j }
  \left\langle Q_{ jk } \right\rangle_{ \text{ch} }
 \bigg]
 \label{eq: S1 operator}
 ,
\end{align}
where the charge mean values are defined by
\begin{align}
 &
 \left\langle r^{ 2 } \right\rangle_{ \text{ch} }
 =
 \frac{ 1 }{ Z }
 \sum_{ a = 1 }^{ Z }
 \big\langle \psi^{( N )}_{ \text{g.s.} } \big|
  r_{ a }^{ 2 }
 \big| \psi^{( N )}_{ \text{g.s.} } \big\rangle
 \\
 &
 \left\langle Q_{ jk } \right\rangle_{ \text{ch} }
 =
 \frac{ 1 }{ Z }
 \sum_{ a = 1 }^{ Z }
 \big\langle \psi^{( N )}_{ \text{g.s.} } \big|
  Q_{ a, jk }
 \big| \psi^{( N )}_{ \text{g.s.} } \big\rangle
 ,
\end{align}
and
\begin{align}
 Q^{( 2 )}_{ a }
 =
 \sqrt{ \frac{ 3 }{ 2 } } \big[ \bm{ r }_{ a } \otimes \bm{ r }_{ a } \big]^{( 2 )}
\end{align}
is the quadrupole moment of proton.
Since the $ P $, $ T $-odd $ \pi NN $ interaction is scalar, only the $ z $-component $ S_{ z } $ can have nonzero values.
The third term of the Schiff moment operator (\ref{eq: S1 operator}) must vanish in spin $ \frac{ 1 }{ 2 } $ nuclei including $^{199}$Hg.

\begin{figure}[htb]
\begin{center}
\includegraphics[width=0.8\linewidth]{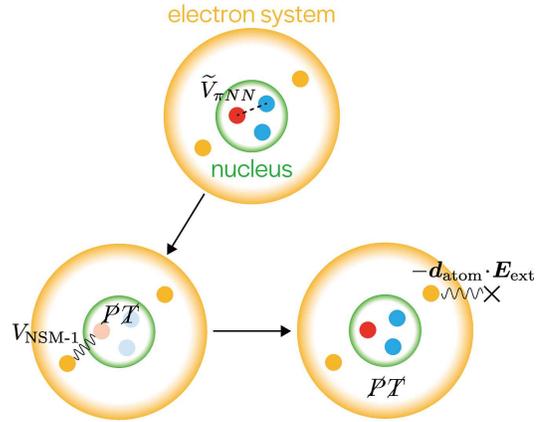}
\caption{
\label{diagram: S1 process to atomic EDM}
Schematic illustration of how the $ P $, $ T $-odd $ \pi NN $ interaction $ \widetilde{ V }_{ \pi NN } $ induces the atomic EDM $ \bm{ d }_{ \text{atom} } $.
The $ P $, $ T $-odd $ \pi NN $ interaction induces the nuclear Schiff moment as well as the nuclear EDM.
The interaction of the nuclear Schiff moment with the electrons $ V_{ \text{NSM-1} } $ violates $ P $ and $ T $ symmetries both in the nucleus and in the electron system.
Finally, the $ P $, $ T $ violation in the electron system generates the atomic EDM.
}
\end{center}
\end{figure}

In conclusion of this section, the leading order contribution from the $ P $, $ T $-odd $ \pi NN $ interaction is given by
\begin{align}
 &
 \Delta E_{ 2 }
 ( \overline{ g }_{ \pi NN }, q_{ N } )
 +
 \Delta E_{ 3 }
 ( \overline{ g }_{ \pi NN }, q_{ N }, q_{ e } )
 \notag\\
 & \quad
 =
 \sum_{ m }
 \frac{ 1 }{ E^{( e )}_{ \text{g.s.} } - E^{( e )}_{ m } }
 \notag\\
 & \qquad \times
 \big\langle \psi^{( e )}_{ \text{g.s.} } \big|
  - \bm{ d }_{ \text{atom} } \! \cdot \! \bm{ E }_{ \text{ext} }
 \big| \psi^{( e )}_{ m } \big\rangle
 \big\langle \psi^{( e )}_{ m } \big|
  V_{ \text{NSM-1} }
 \big| \psi^{( e )}_{ \text{g.s.} } \big\rangle
 \notag\\
 & \qquad
 + c.c.
\end{align}
This result implies that the interaction of the Schiff moment with the electrons denoted by $ V_{ \text{NSM-1} } $ induces the atomic EDM that survives the screening.
The third-order process is illustrated in Fig.~\ref{diagram: S1 process to atomic EDM}.


The Schiff moments $ S_{ 1 } $ of actinide nuclei would be enhanced thanks to octupole correlations and the parity doubling of the ground states~\cite{Ginges2004-review}.
It is expected from recent nuclear many-body calculations~\cite{Engel2003-225Ra-Schiff,Dobaczewski2005,Dobaczewski2018,Yanase2020-129Xe-199Hg} that the Schiff moment of $^{225}$Ra is greater than that of $^{199}$Hg by orders of magnitude, although the uncertainty is still large.

\section{Nucleon EDM \label{sec: S2}}

There are several attempts to identify the leading order contribution from the intrinsic EDMs of nucleons to the atomic EDM.
In particular, the Schiff moment of $^{199}$Hg that originates from the nucleon EDM was computed in the random phase approximation~\cite{Dmitriev2003-PRL}.
Using their result, an upper bound on the neutron EDM was evaluated from the experimental limit on the atomic EDM as $ d_{ n } < 1.6 \times 10^{ -26 } e \text{cm} $~\cite{Graner2016,*Graner2017-erratum}.
This constraint is competitive with a recent direct measurement $ d_{ n } < 1.8 \times 10^{ -26 } e \text{cm} $~\cite{Abel2020}.
On the other hand, it was claimed that the nucleon EDMs in a neutral atom are completely screened~\cite{Oshima2007,Fujita2012}.
In this section, I demonstrate that the screening of the nucleon EDMs is incomplete in the point-like nucleus limit.

Figure~\ref{diagram: nEDM, first order} represents the direct coupling of the nucleon EDMs to the external electric field.
This first-order contribution is given by
\begin{align}
 \Delta E_{ 1 }
 \big( d_{ N } \big)
 =
 \sum_{ a = 1 }^{ A }
 \big\langle \psi^{( N )}_{ \text{g.s.} } \big|
  - \bm{ d }_{ a } \! \cdot \! \bm{ E }_{ \text{ext} }
 \big| \psi^{( N )}_{ \text{g.s.} } \big\rangle
 \label{nEDM int with Eext}
 ,
\end{align}
where $ \bm{ d }_{ a } $ denotes the nucleon EDMs.

The internal interaction of the nucleon EDMs with the electrons
\begin{align}
 \widetilde{ V }^{( e \overline{ N } )}
 & =
 e
 \sum_{ i = 1 }^{ Z }
 \sum_{ a = 1 }^{ A }
 \bm{ d }_{ a }
 \! \cdot \!
 \bm{ \nabla }_{ i }'
 \frac{ 1 }{ \left| \bm{ r }_{ i }' - \bm{ r }_{ a } \right| }
 \label{eq: nucleon EDM int with electron}
\end{align}
violate $ P $ and $ T $ symmetries in the electron system.
Thus, the induced atomic EDM contributes to the energy shift in second order perturbation as
\begin{align}
 &
 \Delta E_{ 2 }
 \big( d_{ N }, q_{ e } \big)
 =
 \sum_{ m }
 \frac{ 1 }{ E^{( e )}_{ \text{g.s.} } - E^{( e )}_{ m } }
 \notag\\
 & \quad \times
 \big\langle \psi^{( A )}_{ \text{g.s.} } \big|
  - \bm{ d }_{ \text{atom} } \! \cdot \! \bm{ E }_{ \text{ext} }
 \big| \psi^{( A )}_{ m } \big\rangle
 \big\langle \psi^{( A )}_{ m } \big|
  \widetilde{ V }^{( e \overline{ N } )}_{ \text{even} - l }
 \big| \psi^{( A )}_{ \text{g.s.} } \big\rangle
 \notag\\
 & \quad
 + c.c.
 \label{eq: Delta E2 ( dn, qe )}
\end{align}
This process is represented in Fig.~\ref{diagram: nuclear EDM from nEDM, second order}.
The internal interaction~(\ref{eq: nucleon EDM int with electron}), which is expanded for $ r_{ i }' > r_{ a } $ as
\begin{align}
 \bm{ d }_{ a }
 \! \cdot \!
 \bm{ \nabla }_{ i }'
 \frac{ 1 }{ \left| \bm{ r }_{ i }' - \bm{ r }_{ a } \right| }
 & =
 \sum_{ l = 0 }^{ \infty }
 \frac{ ( -1 )^{ l } }{ l ! }
 \big( \bm{ r }_{ a } \! \cdot \! \bm{ \nabla }_{ i }' \big)^{ l }
 \bm{ d }_{ a }
 \! \cdot \!
 \bm{ \nabla }_{ i }'
 \frac{ 1 }{ r_{ i }' }
 ,
\end{align}
is restricted to the even-$ l $ components because $ P $ and $ T $ symmetries are not violated in the nuclear system.

The ground state and excited states of the atomic Hamiltonian $ H_{ \text{atom} } $ without $ P $, $ T $-odd interactions are expressed as
\begin{align}
 &
 \big| \psi^{( A )}_{ \text{g.s.} } \big\rangle
 =
 \big| \psi^{( N )}_{ \text{g.s.} } \big\rangle
 \otimes
 \big| \psi^{( e )}_{ \text{g.s.} } \big\rangle
 ,
 \\
 &
 \big| \psi^{( A )}_{ m } \big\rangle
 =
 \big| \psi^{( N )}_{ \text{g.s.} } \big\rangle
 \otimes
 \big| \psi^{( e )}_{ m } \big\rangle
 ,
\end{align}
respectively.


\begin{figure}[htb]
\begin{center}
\subfigure[]{
 \includegraphics[height=4.0cm]{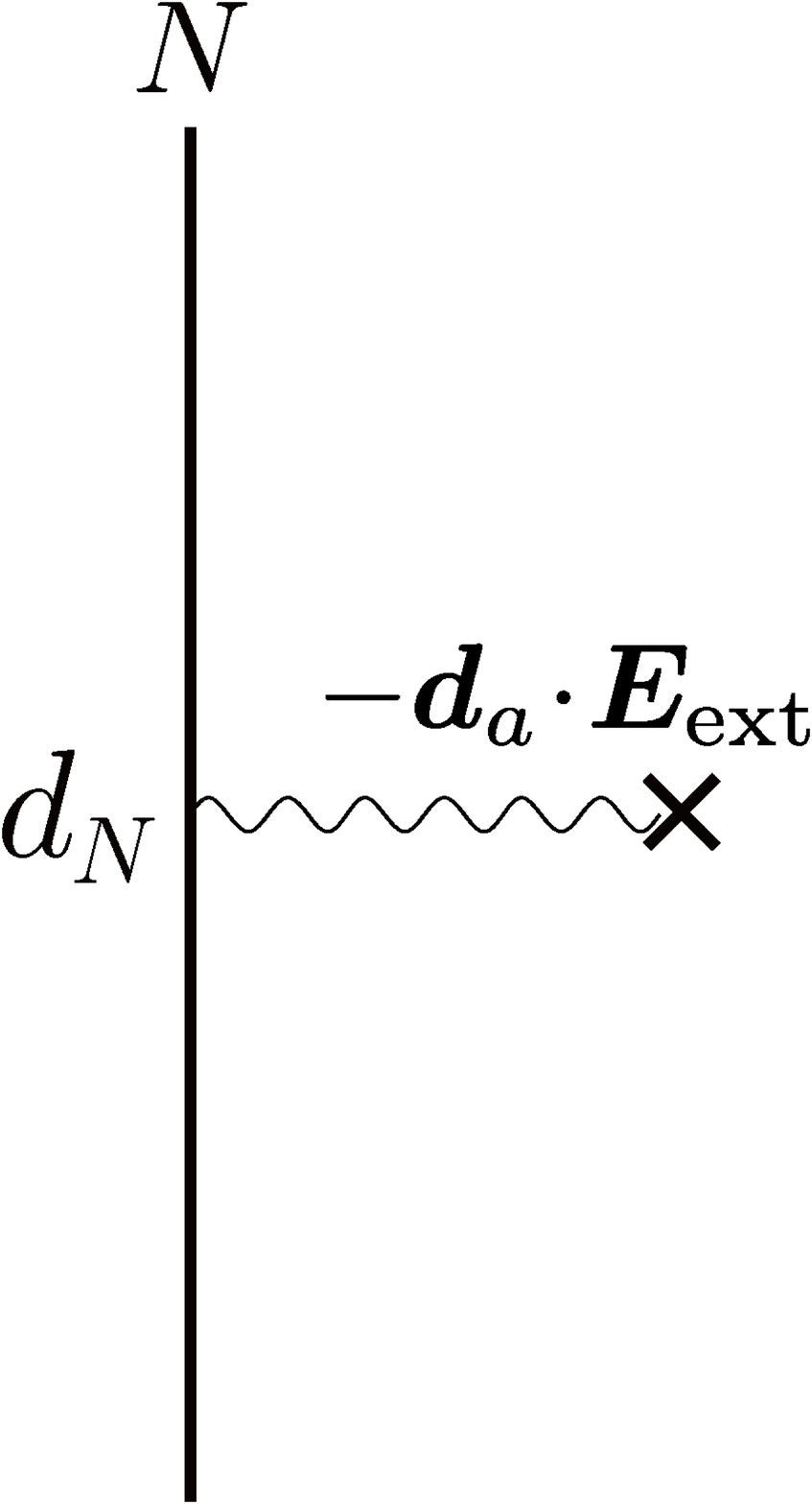}
\label{diagram: nEDM, first order}
}
\hspace{10pt}
\subfigure[]{
 \includegraphics[height=4.0cm]{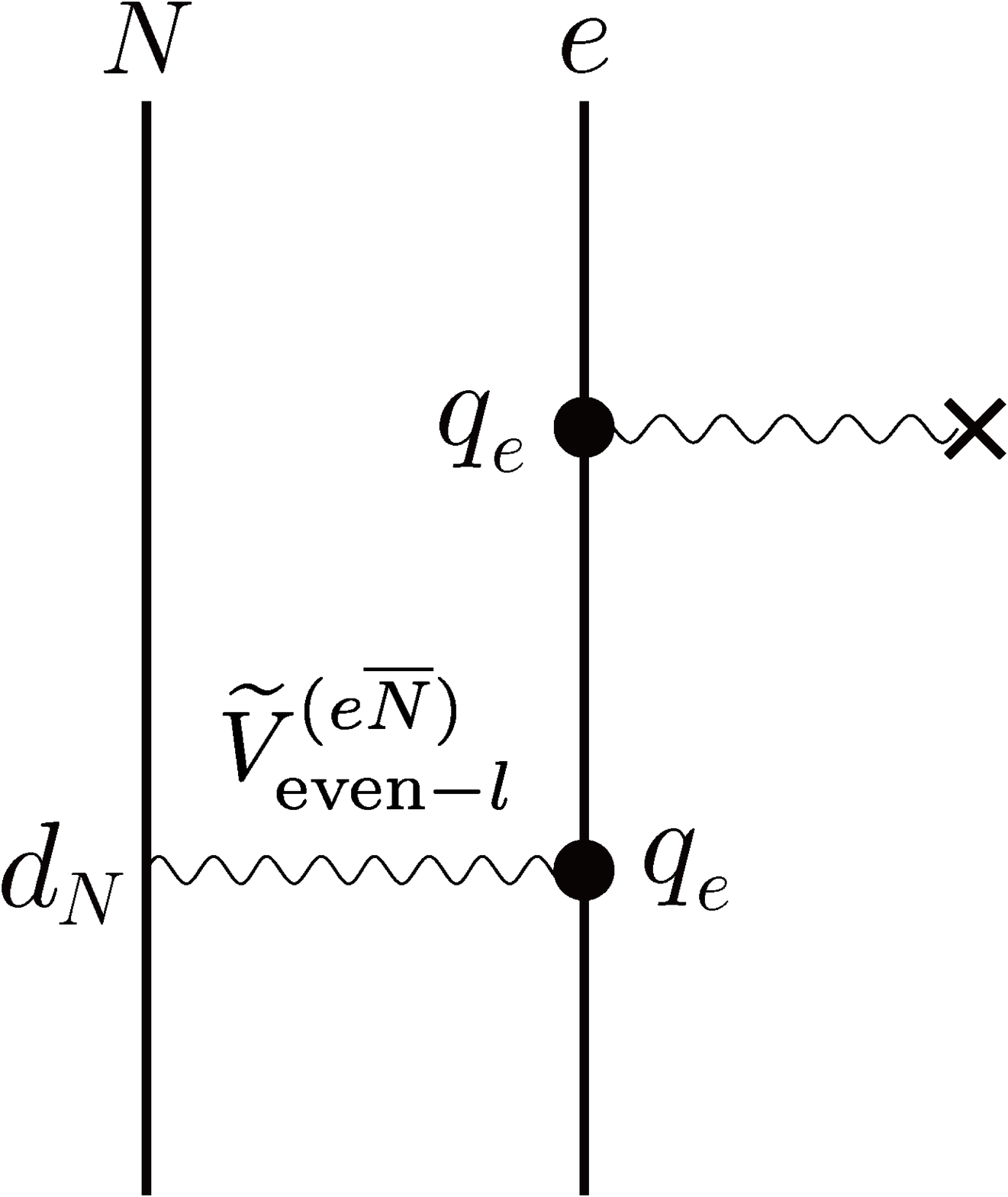}
\label{diagram: nuclear EDM from nEDM, second order}
}
\caption{
\label{diagram: nucleon EDM}
The external interaction of (a) the nucleon EDMs and (b) the atomic EDM induced by the interaction of the nucleon EDMs with the electrons.
The vertices with $ d_{ N } $ indicate the interactions of the nucleon EDM.
}
\end{center}
\end{figure}


I introduce a Hermitian operator
\begin{align}
 U_{ N }
 =
 i
 \frac{ 1 }{ Ze }
 \sum_{ i = 1 }^{ Z }
 \sum_{ a = 1 }^{ A }
 \left\langle \bm{ d }_{ a } \right\rangle
 \! \cdot \!
 \bm{ \nabla }_{ i }'
 \label{eq: Hermitian operator for nucleon EDM}
 ,
\end{align}
where in contrast to $ \langle \bm{ d }_{ \text{nucl} } \rangle $ in Eq.~(\ref{eq: Hermitian operator for nuclear EDM}),
\begin{align}
 \left\langle \bm{ d }_{ a } \right\rangle
 =
 \big\langle \psi^{( N )}_{ \text{g.s.} } \big|
  \bm{ d }_{ a }
 \big| \psi^{( N )}_{ \text{g.s.} } \big\rangle
\end{align}
is the expectation value in the ground state of $ H_{ \text{nucl} } $ conserving $ P $ and $ T $ symmetries.
The external interaction of the nucleon EDMs~(\ref{nEDM int with Eext}) is transformed as
\begin{align}
 &
 \sum_{ a = 1 }^{ A }
 \big\langle \psi^{( A )}_{ \text{g.s.} } \big|
  -
  \bm{ d }_{ a }
  \! \cdot \!
  \bm{ E }_{ \text{ext} }
 \big| \psi^{( A )}_{ \text{g.s.} } \big\rangle
 \notag\\
 & \quad
 =
 i
 \big\langle \psi^{( A )}_{ \text{g.s.} } \big|
  \big[
   U_{ N },
   -
   \bm{ d }_{ \text{atom} }
   \! \cdot \!
   \bm{ E }_{ \text{ext} }
  \big]
 \big| \psi^{( A )}_{ \text{g.s.} } \big\rangle
 .
\end{align}
The $ l = 0 $ component of the internal interaction~(\ref{eq: nucleon EDM int with electron}), which is explicitly given by
\begin{align}
 \widetilde{ V }^{( e \overline{ N } )}_{ l = 0 }
 =
 e
 \sum_{ i = 1 }^{ Z }
 \sum_{ a = 1 }^{ A }
 \bm{ d }_{ a }
 \! \cdot \!
 \bm{ \nabla }_{ i }'
 \frac{ 1 }{ r_{ i }' }
 ,
\end{align}
is transformed as
\begin{align}
 &
 \big\langle \psi^{( A )}_{ m } \big|
  \widetilde{ V }^{( e \overline{ N } )}_{ l = 0 }
 \big| \psi^{( A )}_{ \text{g.s.} } \big\rangle
 \notag\\
 & \quad
 =
 i
 \big\langle \psi^{( A )}_{ m } \big|
  \big[
   U_{ N },
   V^{( eN )}_{ l = 0 }
  \big]
 \big| \psi^{( A )}_{ \text{g.s.} } \big\rangle
 \notag\\
 & \quad
 =
 i
 \big\langle \psi^{( A )}_{ m } \big|
  \big[
   U_{ N },
   H_{ e }
  \big]
 \big| \psi^{( A )}_{ \text{g.s.} } \big\rangle
 \notag\\
 & \qquad
 -
 i
 \big\langle \psi^{( A )}_{ m } \big|
  \big[
   U_{ N },
   V^{( eN )}_{ l = 2 }
   +
   V^{( eN )}_{ l = 4 }
   + \cdots
  \big]
 \big| \psi^{( A )}_{ \text{g.s.} } \big\rangle
 \label{eq: U_N trans of nucleon EDM int, l = 0}
 ,
\end{align}
where $ [ U_{ N }, T_{ e } ] = 0 $ and $ [ U_{ N }, V^{( ee )} ] = 0 $ are used.
Substituting (\ref{eq: U_N trans of nucleon EDM int, l = 0}) into (\ref{eq: Delta E2 ( dn, qe )}), one can find
\begin{align}
 &
 \Delta E_{ 1 }
 \big( d_{ N } \big)
 +
 \Delta E_{ 2 }
 \big( d_{ N }, q_{ e } \big)
 \notag\\
 & \quad
 =
 \sum_{ m }
 \frac{ 1 }{ E^{( e )}_{ \text{g.s.} } - E^{( e )}_{ m } }
 \notag\\
 & \qquad \times
 \big\langle \psi^{( e )}_{ \text{g.s.} } \big|
  -
  \bm{ d }_{ \text{atom} }
  \! \cdot \!
  \bm{ E }_{ \text{ext} }
 \big| \psi^{( e )}_{ m } \big\rangle
 \big\langle \psi^{( e )}_{ m } \big|
  \widetilde{ V }_{ \text{NSM-2} }
 \big| \psi^{( e )}_{ \text{g.s.} } \big\rangle
 \notag\\
 & \qquad
 + c.c.
 \label{eq: Delta E1 ( dn ) + Delta E2 ( dn, qe )}
\end{align}
The right-hand side vanishes for a point-like nucleus, where the $ eN $ interactions in Eqs.~(\ref{eq: charge int}) and (\ref{eq: nucleon EDM int with electron}) are restricted to $ l \leq 1 $.
The complete screening of a point-like nucleus is valid even if the nucleons are relativistic~\cite{Liu2007-nEDM}.


\begin{figure}[htb]
\begin{center}
\includegraphics[width=0.8\linewidth]{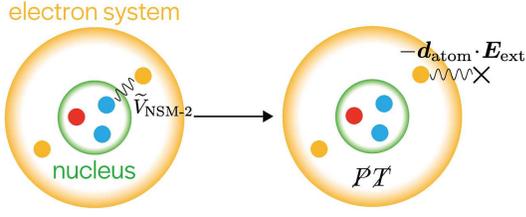}
\caption{
\label{diagram: S2 process to atomic EDM}
The leading order contribution of the nucleon EDMs to the atomic EDM $ \bm{ d }_{ \text{atom} } $.
The $ P $, $ T $-odd $ eN $ interaction due to the finite-size effect $ \widetilde{ V }_{ \text{NSM-2} } $ appears in the point-like nucleus limit.
}
\end{center}
\end{figure}


The remaining second-order process in Eq.~(\ref{eq: Delta E1 ( dn ) + Delta E2 ( dn, qe )}) is illustrated in Fig.~\ref{diagram: S2 process to atomic EDM}.
In the point-like nucleus limit, the finite-size effect is given up to $ l = 2 $ by
\begin{align}
 &
 \big\langle \psi^{( e )}_{ m } \big|
  \widetilde{ V }_{ \text{NSM-2} }
 \big| \psi^{( e )}_{ \text{g.s.} } \big\rangle
 \notag\\
 & \quad
 =
 \big\langle \psi^{( A )}_{ m } \big|
  \widetilde{ V }^{( e \overline{ N } )}_{ l = 2 }
 \big| \psi^{( A )}_{ \text{g.s.} } \big\rangle
 - i
 \big\langle \psi^{( A )}_{ m } \big|
  \Big[
   U_{ N } , V^{( eN )}_{ l = 2 }
  \Big]
 \big| \psi^{( A )}_{ \text{g.s.} } \big\rangle
 \notag\\
 & \quad
 =
 \big\langle \psi^{( e )}_{ m } \big|
  - 4 \pi e
  \sum_{ i = 1 }^{ Z }
  \big\langle \bm{ S }_{ 2 } \big\rangle
  \! \cdot \!
  \bm{ \nabla }_{ i }' \delta ( \bm{ r }_{ i }' )
 \big| \psi^{( e )}_{ \text{g.s.} } \big\rangle
 \label{eq: NSM-2 int}
 ,
\end{align}
as derived in Appendix~\ref{app: nEDM int}.
The nuclear moment $ \bm{ S }_{ 2 } $ is also called the Schiff moment, and given by
\begin{align}
 S_{ 2, k }
 & =
 \frac{ 1 }{ 6 }
 \sum_{ a = 1 }^{ A }
 d_{ a, k }
 \big(
  r_{ a }^{ 2 }
  -
  \big\langle r^{ 2 } \big\rangle_{ \text{ch} }
 \big)
 \notag\\
 & \quad
 +
 \frac{ 2 }{ 15 }
 \sum_{ a = 1 }^{ A }
 d_{ a, j }
 \big(
  Q_{ a, jk }
  -
  \big\langle Q_{ jk } \big\rangle_{ \text{ch} }
 \big)
 \label{eq: S2 operator}
 .
\end{align}

Using the independent particle model (IPM)~\cite{Yanase2020-129Xe-199Hg}, one obtains
\begin{align}
 S_{ 2 } \left( ^{199} \text{Hg} \right)
 =
 2.8 d_{ n } \, ( \text{fm}^{ 2 } )
 \label{eq: S2 value in IPM}
 ,
\end{align}
which is consistent with the previous evaluation~\cite{Ginges2004-review} $ S_{ 2 } \simeq 2.2 d_{ n } \, ( \text{fm}^{ 2 } ) $.
This quantity was calculated as $ S_{ 2 } = ( 1.895 \pm 0.035 ) d_{ n } \, ( \text{fm}^{ 2 } )$ in the random phase approximation~\cite{Dmitriev2003-PRL}.

\section{Next-to-leading order contribution of nucleon EDM \label{sec: S3}}

As discussed in Sec.~\ref{sec: S2}, the atomic EDM is sensitive to the Schiff moment $ S_{ 2 } $, which stems from the screening effect of the nucleon EDMs themselves.
In addition to the nucleon EDMs, the nuclear EDM is independently coupled to the external electric field as shown in Eq.~(\ref{eq: V_ext}).
The nuclear EDM is induced not only by the $ \pi NN $ interaction but also by the interaction between the nucleon EDMs and the protons
\begin{align}
 \widetilde{ V }^{( N \overline{ N } )}
 & =
 e d_{ p }
 \sum_{ a \neq b }^{ Z }
 \frac{
  \bm{ \sigma }_{ a }
  \! \cdot \!
  ( \bm{ r }_{ b } - \bm{ r }_{ a } )
 }
 { \left| \bm{ r }_{ b } - \bm{ r }_{ a } \right|^{ 3 } }
 \notag\\
 & \quad
 +
 e d_{ n }
 \sum_{ b = 1 }^{ Z }
 \sum_{ a = 1 }^{ N }
 \frac{
  \bm{ \sigma }_{ a }
  \! \cdot \!
  ( \bm{ r }_{ b } - \bm{ r }_{ a } )
 }
 { \left| \bm{ r }_{ b } - \bm{ r }_{ a } \right|^{ 3 } }
 .
\end{align}


\begin{figure}[htb]
\begin{center}
\subfigure[]{
 \includegraphics[height=3.5cm]{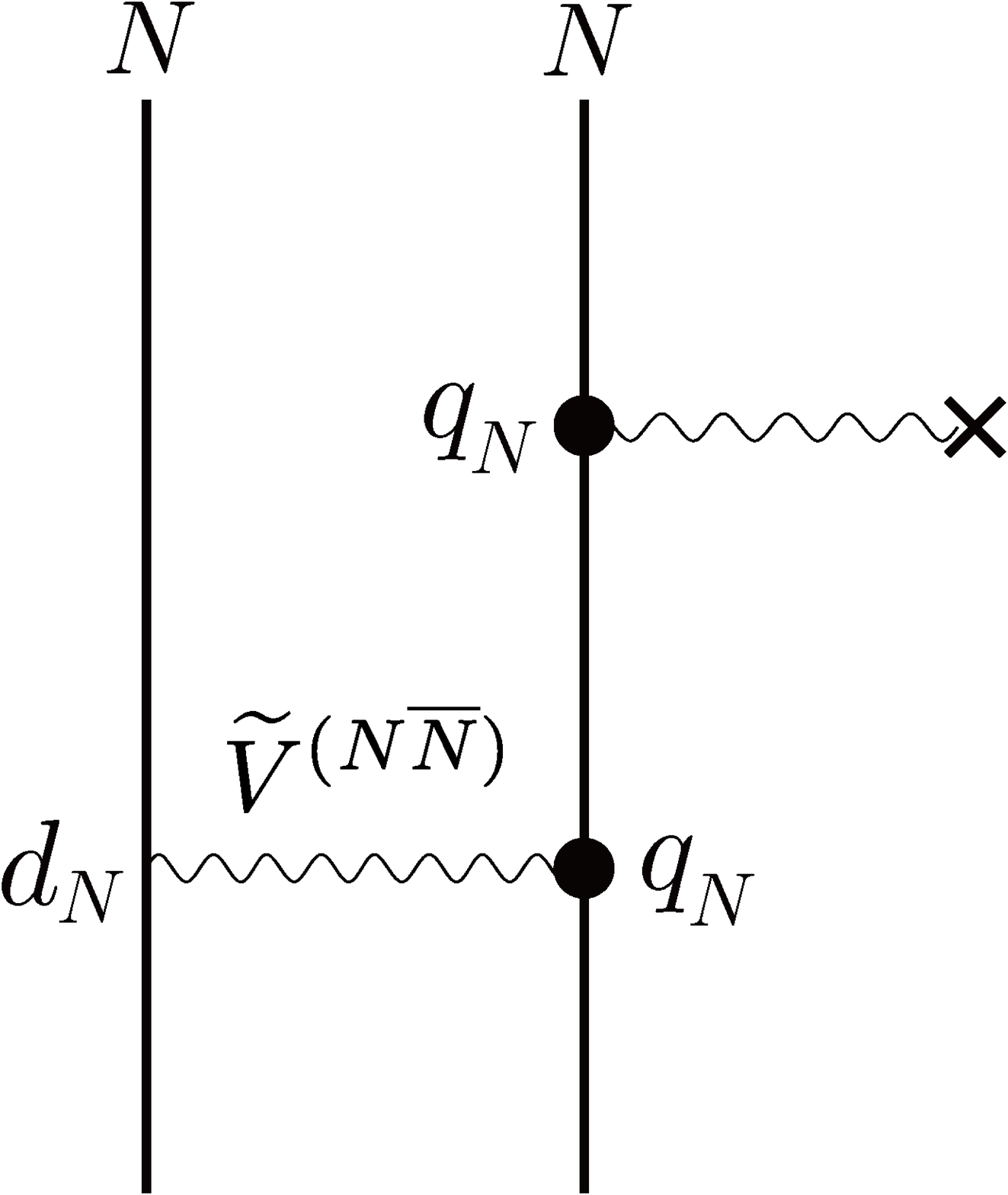}
\label{diagram: Schiff moment from nEDM, second order}
}
\hspace{10pt}
\subfigure[]{
 \includegraphics[height=3.5cm]{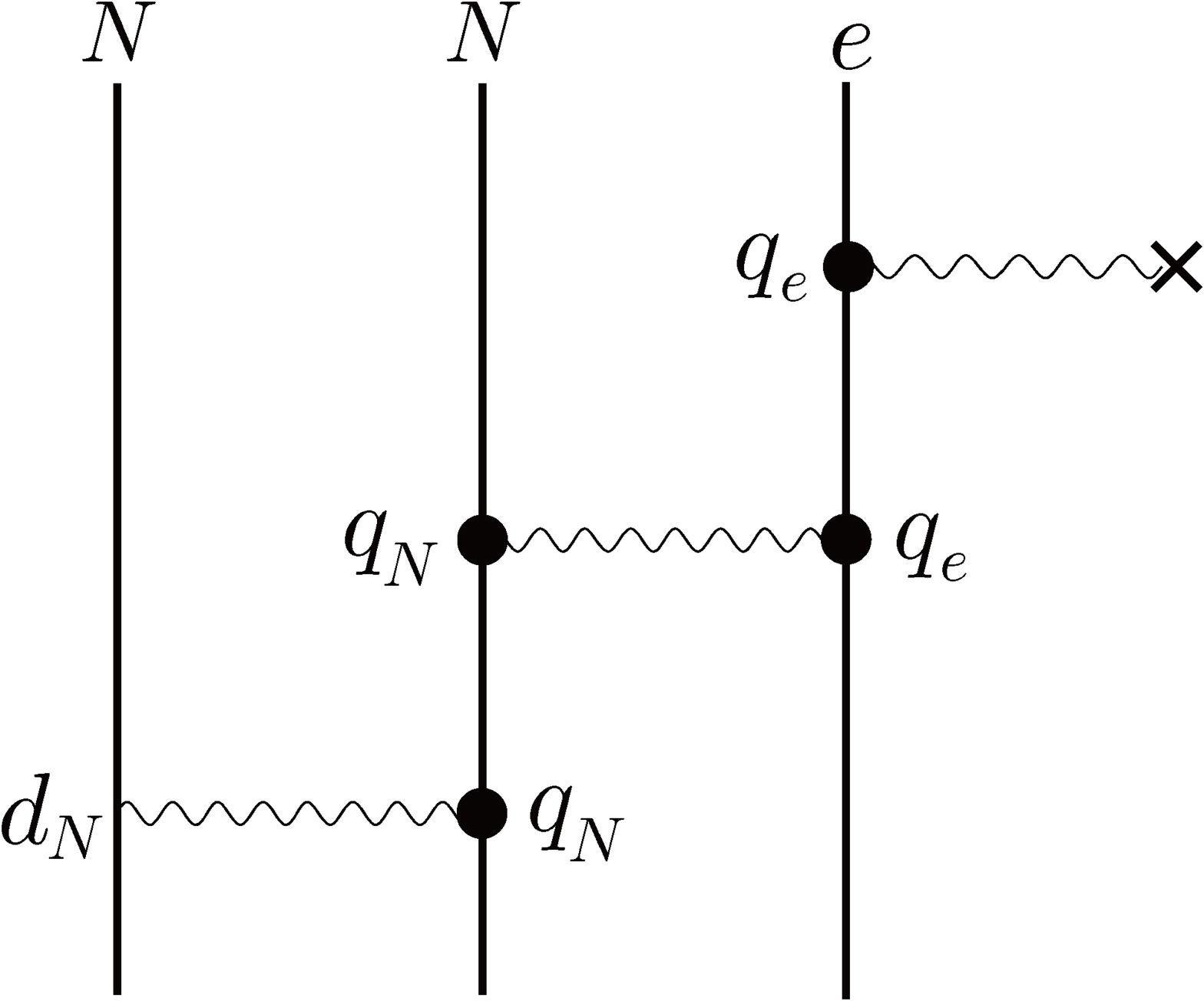}
\label{diagram: Schiff moment from nEDM, third order}
}
\caption{
\label{diagram: Schiff moment from nEDM}
(a) The second-order and (b) the third-order contributions of the interactions between the nucleon EDMs and the protons.
}
\end{center}
\end{figure}


A similar argument as in Sec.~\ref{sec: S1} shows that this contribution represented in Fig.~\ref{diagram: Schiff moment from nEDM, second order} is screened by the third-order processes represented in Fig.~\ref{diagram: Schiff moment from nEDM, third order}.
The finite-size effect leads to the next-to-leading order contribution of the nucleon EDM to the Schiff moment
\begin{align}
 &
 \big\langle \bm{ S }_{ 3 } \big\rangle
 =
 \sum_{ n }
 \frac{ 1 }{ E^{( N )}_{ \text{g.s.} } - E^{( N )}_{ n } }
 \notag\\
 & \qquad \qquad
 \times
 \big\langle \psi^{( N )}_{ \text{g.s.} } \big|
  \bm{ S }_{ 3 }
 \big| \psi^{( N )}_{ n } \big\rangle
 \big\langle \psi^{( N )}_{ n } \big|
  \widetilde{ V }^{( N \overline{ N } )}
 \big| \psi^{( N )}_{ \text{g.s.} } \big\rangle
 \notag\\
 & \qquad \quad
 + c.c.
 ,
\end{align}
where the operator $ \bm{ S }_{ 3 } $ is the same as $ \bm{ S }_{ 1 } $.
This correction is evaluated as
\begin{align}
 S_{ 3 } \left( ^{199} \text{Hg} \right)
 =
 - 0.15 d_{ n } \, ( \text{fm}^{ 2 } )
\end{align}
in the IPM.

\section{Conclusion \label{sec: conclusion}}

I have examined the screening of the intrinsic EDMs of nucleons and the nuclear EDM in a neutral atom.
In the point-like nucleus limit, the Schiff moment of a finite-size nucleus induces the atomic EDM that circumvents the screening.
The total Schiff moment is given by
\begin{align}
 S = S_{ 1 } + S_{ 2 } + S_{ 3 }
 ,
\end{align}
where $ S_{ 2 } $ and $ S_{ 3 } $ are due to the nucleon EDM.
The nucleon EDM contributions provide constraints on the short-range component, whereas the $ P $, $ T $-odd $ \pi NN $ interaction contributes to the nucleon EDM in the leading order chiral perturbation theory~\cite{Crewther1979,*Crewther1980-erratum,Yamanaka2017-review,Chupp2019-review}.

The leading order contribution $ S_{ 2 } $ stems from the screening of the nucleon EDMs themselves.
The nuclear EDM is induced by the interaction of the nucleon EDMs with the protons as well as the $ P $, $ T $-odd $ \pi NN $ interaction.
The screening of the nuclear EDM gives rise to the next-to-leading order contribution to the Schiff moment $ S_{ 3 } $.
This correction to the Schiff moment of $^{199}$Hg is of the order of $ 5 \% $ in the IPM.
Here, nuclear octupole correlations would enhance $ S_{ 3 } $ as well as $ S_{ 1 } $, which is induced by the $ \pi NN $ interaction, by orders of magnitude.
Consequently, the dependence of the Schiff moment on the nucleon EDM can be dominated by $ S_{ 3 } $ rather than $ S_{ 2 } $ in octupole deformed nuclei.

\appendix
\section{Schiff moment due to the $ P $, $ T $-odd $ \pi NN $ interaction \label{app: Schiff from pi-NN}}

The Schiff moment operator $ \bm{ S }_{ 1 } $ is defined by the matrix elements of the remaining $ eN $ interaction
\begin{align}
 &
 V^{( eN )}_{ l = 3 }
 - i
 \big[
  U_{ \text{nucl} } , V^{( eN )}_{ l = 2 }
 \big]
 \notag\\
 & \quad
 =
 \frac{ 1 }{ 6 } e^{ 2 }
 \sum_{ i = 1 }^{ Z }
 \sum_{ a = 1 }^{ Z }
 \big(
  \bm{ r }_{ a } \! \cdot \! \bm{ \nabla }_{ i }'
 \big)^{ 3 }
 \frac{ 1 }{ r_{ i }' }
 \notag\\
 & \qquad
 -
 \frac{ e }{ 2Z }
 \sum_{ i = 1 }^{ Z }
 \sum_{ a = 1 }^{ Z }
 \big(
  \bm{ r }_{ a } \! \cdot \! \bm{ \nabla }_{ i }'
 \big)^{ 2 }
 \left\langle \bm{ d }_{ \text{nucl} } \right\rangle
 \! \cdot \!
 \bm{ \nabla }_{ i }'
 \frac{ 1 }{ r_{ i }' }
 .
 \label{eq: Schiff moment int, bare}
\end{align}
The nuclear part can be separated as
\begin{align}
 \big(
  \bm{ r }_{ a } \! \cdot \! \bm{ \nabla }_{ i }'
 \big)^{ 3 }
 =
 \frac{ 3 }{ 5 }
 r_{ a }^{ 2 }
 \big(
  \bm{ r }_{ a } \! \cdot \! \bm{ \nabla }_{ i }'
 \big)
 \bm{ \nabla }_{ i }'^{ 2 }
 +
 \frac{ 2 }{ 5 }
 Q_{ a }^{( 3 )} \! \cdot \! \nabla_{ i }'^{( 3 )}
 ,
 \label{eq: tensor products r.nabla^3}
\end{align}
and
\begin{align}
 &
 \big(
  \bm{ r }_{ a } \! \cdot \! \bm{ \nabla }_{ i }'
 \big)^{ 2 }
 \left\langle \bm{ d }_{ \text{nucl} } \right\rangle
 \! \cdot \!
 \bm{ \nabla }_{ i }'
 \notag\\
 & \quad
 =
 \frac{ 1 }{ 3 }
 r_{ a }^{ 2 }
 \left\langle \bm{ d }_{ \text{nucl} } \right\rangle
 \! \cdot \!
 \bm{ \nabla }_{ i }'
 \bm{ \nabla }_{ i }'^{ 2 }
 \notag\\
 & \qquad
 -
 \frac{ 2 }{ 3 }
 \sqrt{ \frac{ 2 }{ 5 } }
 \Big[
  \left\langle \bm{ d }_{ \text{nucl} } \right\rangle
  \otimes
  Q_{ a }^{( 2 )}
 \Big]^{( 1 )}
 \! \cdot \!
 \bm{ \nabla }_{ i }'
 \bm{ \nabla }_{ i }'^{ 2 }
 \notag\\
 & \qquad
 -
 \frac{ 2 }{ \sqrt{ 15 } }
 \Big[
  \left\langle \bm{ d }_{ \text{nucl} } \right\rangle
  \otimes
  Q_{ a }^{( 2 )}
 \Big]^{( 3 )}
 \! \cdot \!
 \nabla_{ i }'^{( 3 )}
 ,
 \label{eq: tensor products r.nabla^2 d.nabla}
\end{align}
where
\begin{align}
 Q_{ a }^{( 3 )}
 =
 \sqrt{ \frac{ 5 }{ 2 } }
 \Big[
  \big[
   \bm{ r }_{ a } \otimes \bm{ r }_{ a }
  \big]^{( 2 )}
  \otimes
  \bm{ r }_{ a }
 \Big]^{( 3 )}
\end{align}
is the nuclear octupole moment and
\begin{align}
 \nabla_{ i }'^{( 3 )}
 =
 \sqrt{ \frac{ 5 }{ 2 } }
 \Big[
  \big[
   \bm{ \nabla }_{ i }' \otimes \bm{ \nabla }_{ i }'
  \big]^{( 2 )}
  \otimes
  \bm{ \nabla }_{ i }'
 \Big]^{( 3 )}
\end{align}
is a rank 3 operator of electron.
Since the last terms in Eqs.~(\ref{eq: tensor products r.nabla^3}) and (\ref{eq: tensor products r.nabla^2 d.nabla}) can be omitted~\cite{Ginges2004-review}, Eq.~(\ref{eq: Schiff moment int, bare}) is rewritten as
\begin{align}
 &
 V^{( eN )}_{ l = 3 }
 - i
 \big[
  U_{ \text{nucl} } , V^{( eN )}_{ l = 2 }
 \big]
 \notag\\
 & \quad
 =
 \frac{ 1 }{ 10 } e
 \sum_{ i = 1 }^{ Z }
 \sum_{ a = 1 }^{ Z }
 \bigg[
  e
  r_{ a }^{ 2 } \bm{ r }_{ a }
 -
 \frac{ 5 }{ 3 Z }
 r_{ a }^{ 2 }
 \left\langle \bm{ d }_{ \text{nucl} } \right\rangle
 \notag\\
 & \qquad
 +
 \frac{ 2 }{ 3 Z }
 \sqrt{ 10 }
 \Big[
  \left\langle \bm{ d }_{ \text{nucl} } \right\rangle
  \otimes
  Q_{ a }^{( 2 )}
 \Big]^{( 1 )}
 \bigg]
  \! \cdot \! \bm{ \nabla }_{ i }'
  \bm{ \nabla }_{ i }'^{ 2 }
  \frac{ 1 }{ r_{ i }' }
 \label{eq: Schiff moment int, without octupole}
 .
\end{align}
In the point-like nucleus limit, $ R \rightarrow 0 $, one then obtain the Schiff moment interaction $ V_{ \text{NSM-1} } $ in Eq.~(\ref{eq: NSM-1 int}).

\section{Leading order contribution of nucleon EDM \label{app: nEDM int}}

The $ P $, $ T $-odd interactions between the nucleon EDMs and the electrons in Eq.~(\ref{eq: Delta E1 ( dn ) + Delta E2 ( dn, qe )}) are written as
\begin{align}
 &
 \widetilde{ V }_{ l = 2 }^{( e \overline{ N } )}
 - i
 \left[
  U_{ N },
  V_{ l = 2 }^{( eN )}
 \right]
 \notag\\
 & \quad
 =
 \frac{ 1 }{ 2 } e
 \sum_{ i = 1 }^{ Z }
 \sum_{ a = 1 }^{ A }
  ( \bm{ r }_{ a } \! \cdot \! \bm{ \nabla }_{ i }' )^{ 2 }
  \bm{ d }_{ a }
  \! \cdot \!
  \bm{ \nabla }_{ i }'
  \frac{ 1 }{ r_{ i }' }
 \notag\\
 & \qquad
 -
 \frac{ e }{ 2 }
 \sum_{ i = 1 }^{ Z }
 \sum_{ a = 1 }^{ Z }
  ( \bm{ r }_{ a } \! \cdot \! \bm{ \nabla }_{ i }' )^{ 2 }
  \left\langle \bm{ d }_{ N } \right\rangle_{ \text{ch} }
  \! \cdot \!
  \bm{ \nabla }'
  \frac{ 1 }{ r_{ i }' }
 \label{eq: NSM-2, unseparated}
 ,
\end{align}
where
\begin{align}
 \left\langle \bm{ d }_{ N } \right\rangle_{ \text{ch} }
 =
 \frac{ 1 }{ Z }
 \sum_{ a = 1 }^{ A }
 \big\langle \bm{ d }_{ a } \big\rangle
 .
\end{align}
The nuclear part can be separated by using Eq.~(\ref{eq: tensor products r.nabla^2 d.nabla}).
In the point-like nucleus limit, one obtains the Schiff moment interaction $ \widetilde{ V }_{ \text{NSM-2} } $ in Eq.~(\ref{eq: NSM-2 int}).

\begin{acknowledgements}
This research was supported by MEXT as ``Program for Promoting Researches on the Supercomputer Fugaku'' (Simulation for basic science: from fundamental laws of particles to creation of nuclei) and JICFuS.
I used the shell-model code KSHELL~\cite{Shimizu2019-KSHELL} to obtain the nuclear wave function of $^{199}$Hg in the IPM.
I acknowledge Noritaka Shimizu for helpful discussions.
\end{acknowledgements}

\providecommand{\noopsort}[1]{}\providecommand{\singleletter}[1]{#1}

\end{document}